\documentclass[10pt]{article}

\usepackage{amsmath}
\usepackage{amsfonts}

\usepackage{graphicx}

\begin{document}

\title{$p$-Adic numbers in bioinformatics: from genetic code to PAM-matrix}

\author{A.Yu.Khrennikov\footnote{International Center for Mathematical
Modelling in Physics and Cognitive Sciences, University of
V\"axj\"o, S-35195, Sweden, e--mail: Andrei.Khrennikov@vxu.se},
S.V.Kozyrev\footnote{Steklov Mathematical Institute, Moscow, Russia,
e--mail: kozyrev@mi.ras.ru}}

\maketitle

\begin{abstract}
In this paper we denonstrate that the use of the system of 2-adic
numbers provides a new insight to some problems of genetics, in
particular, generacy of the genetic code and the structure of the
PAM matrix in bioinformatics.  The 2-adic distance is an ultrametric
and applications of ultrametrics in bioinformatics are not
surprising. However, by using the 2-adic numbers we match
ultrametric with a number theoretic structure. In this way we find
new applications of an ultrametric which differ from known up to now
in bioinformatics.

We obtain the following results.  We show that the PAM matrix $A$
allows the expansion into the sum of the two matrices
$A=A^{(2)}+A^{(\infty)}$, where the matrix $A^{(2)}$ is 2--adically
regular (i.e. matrix elements of this matrix are close to locally
constant with respect to the discussed earlier by the authors
2--adic parametrization of the genetic code), and the matrix
$A^{(\infty)}$ is sparse. We discuss the structure of the matrix
$A^{(\infty)}$ in relation to the side chain properties of the
corresponding amino acids.
\end{abstract}

\section{Introduction}

Various clustering procedures play a crucial role in bioinformatics,
in particular, genetics, see, e.g.,  \cite{PAM,Isaev} or \cite{TTT}.
 An important
class of such procedures is based on introduction of various metrics
on the space information strings, see e.g. \cite{Murtagh}. A metric
with new interesting features was recently used in theoretical
physics (from string theory to theory of disordered systems, spin
glasses), see e.g. \cite{VVZ}, \cite{Andr}, \cite{Kozyrev},
\cite{obzor} in cognitive science, psychology and image analysis
\cite{Andr3}. This is so called $p$--adic metric (in fact, a class
of metrics depending on the parameter $p$
--- a prime number). The main distinguishing feature of this metric
is its sensitivity to hierarchic patterns in information  having a
special structure matching with $p$--adic encoding of information.\footnote{We remark that to appeal to hierarchical
structures is quite common in genetics and bioinformatics in general, see,
e.g. \cite{TTT}, \cite{TTT1}. Our main contribution is combining the hierarchic structure
approach with number theory.}

A few years ago 2--adic metric was applied to study the problem of
degeneration of the genetic code, see \cite{genetic_code,DD,Andr5}.
These $p$--adic models can be considered as new development in the
approach to investigation of the structure of the genetic code from
the point of view of coding theory, see \cite{Swanson, SjoWold,
Pattern}.

In the present paper we discuss the structure of the PAM matrix used
in bioinformatics (see for example \cite{PAM}) from the point of
view of $p$--adic analysis. We use the 2--adic parametrization of
the genetic (amino acid) code obtained in \cite{genetic_code} (see
also \cite{DD} for the different $p$--adic parametrization).

In \cite{genetic_code,DD} it was shown that, after some special
parametrization of the space of codons (triples of nucleotides) the
genetic code becomes a locally constant map of $p$--adic argument.
Moreover, the degeneracy of the genetic code in this language takes
the form of local constancy of the corresponding mapping.

Let us also mention the application of the $p$--adic parametrization
to the description of the Parisi matrix from the replica symmetry
breaking approach to spin glasses \cite{ABK,PaSu}. After the
$p$--adic parametrization of the numbers of the lines and the
columns the Parisi matrix becomes a locally constant block matrix.

It is natural to check, using the $p$--adic parametrization
approach, the structure of the PAM matrix. The PAM matrix is used in
bioinformatics for sequence alignment and is constructed using a
Markov chain model of point mutations for a protein chain.

We assume that the structure of the PAM matrix has some relation to
the structure of the genetic code. Using this idea we enumerate the
lines and the columns of the PAM matrix using the 2--adic
parametrization of the genetic code. After this parametrization the
PAM matrix becomes more regular, namely, the dependence of the
matrix elements $A_{ij}$ of the PAM matrix on the indices $i$ and
$j$ is close to locally constant with respect to the 2--adic norm
for the majority of matrix elements.

We have some exceptions from this rule. It is easy to see that these
exceptions are related to several amino acids, namely to Y, W, C, F,
L. In order to describe this deviations from 2--adicity we introduce
the following construction: we expand (by hands) the PAM matrix into
the sum of the two matrices
$$
A=A^{(2)}+A^{(\infty)}.
$$
The matrix in this expansion $A^{(2)}$ is 2--adically regular (close
to locally constant). The matrix $A^{(\infty)}$ is sparse (the
majority of matrix elements are zero, non zero matrix elements are
mainly concentrated of the lines and columns related to the amino
acids Y, W, C, F, L).

One can see that the deviations from 2--adicity (i.e. non--zero
matrix elements of $A^{(\infty)}$) are related to amino acids which
are in some sense special --- to the aromatic amino acids Y, W, F,
and to Cysteine C which contains the SH group.

We also mention that the 2--adic structure of the genetic code is
related to some chemical properties of the amino acids. In
particular, hydrophobic amino acids are clustered in two ball with
respect to the 2--adic norm. Therefore the 2--adic parametrization
allows to separate the impact of the chemical and geometrical
properties of aromatic amino acids for the structure of the PAM
matrix.

The structure of the present paper is the following.

In Section 2 we discuss some family of ultrametric spaces.

In Section 3 we describe the 2--adic 2--dimensional parametrization
of the genetic code of \cite{genetic_code}.

In Section 4 we put the PAM250 matrix.

In Section 5 we describe the reshuffling of the lines and the
columns of the PAM matrix, corresponding to the 2--adic
parametrization of the genetic code of Section 2.

In Section 6 we introduce the expansion of the PAM matrix into the
sum of the two matrices, one of which is 2--adically regular (close
to locally constant) and the other is sparse (majority of matrix
elements are equal to zero).

Sections 7 and 8 are appendices where the definitions of PAM
matrices and the eucaryotic genetic code are exposed.

\section{Ultrametric spaces}

An {\it ultrametric space} is a metric space where the metric
$d(x,y)$ satisfies the strong triangle inequality:
$$
d(x,y)\le {\rm max}\left(d(x,z),d(y,z)\right),\qquad \forall x,y,z.
$$
The strong triangle inequality can be stated geometrically: {\it
each side of a triangle is at most as long as the longest one of the
two other sides.} Such  a triangle is quite restricted when
considered in the ordinary Euclidean space --- it is {\it
isosceles,} i.e., $d(x,y) = d(y,z)$ or $d(x,z) = d(y,z)$ or $d(x,y)
= d(z,x).$

An ultrametric space is a natural mathematical object for
description of a {\it hierarchical system.} On ultrametric spaces
there exist many locally constant functions, i.e., functions which
are constant on some vicinity of any point, but not necessarily
constant on the whole space.  In particular, we show that the
genetic code can be considered as a locally constant map on a
specially designed ultrametric space, so called 2-adic plane, see
\cite{genetic_code}.

Let $(X, d)$ be an ultrametric space. We consider  balls $U_r(a)= \{
x \in X:  d(x,a)\leq r\}, r>0, x \in X.$ So, $a$ is the center of
the ball  $U_r(a)$ having radius $r.$ We mention a few unusual (from
the viewpoint of usual Euclidean geometry) properties of ultrametric
balls:

\medskip

a).   Each point of $U_r(a)$ can be chosen as its center. So, inside
a ball all points have ''equal rights''.

\medskip

b).  Any two balls either do not intersect or one of the balls
contains the other ball. In this framework, clustering into {\it
disjoint balls} is a very natural operation.

\medskip

We remark that ultrametric spaces were widely used in
bioinformatics, see \cite{PAM,Isaev}. However, in this paper we plan
to elaborate new applications of ultrametric spaces to biology
(genetics) which are different from mentioned ones.

We are interested in the following special class of ultrametric
spaces $(X,d).$ Every point $x$ is the infinite sequence of digits
\begin{equation}
\label{K} x=(\alpha_0,\alpha_1,\dots,\alpha_n,\dots)\;.
\end{equation}
Each digit yields the finite number of values,
\begin{equation}
\label{K1} \alpha_i\in A_m=\{ 0,\dots,m-1\},
\end{equation}
where $m > 1$ is a natural number, the base of the alphabet $A_m$.

If the sequence $x=(\alpha_0,\alpha_1,\dots,\alpha_n,\dots)$
contains only finite number of non-zero terms
$(\alpha_0,\alpha_1,\dots,\alpha_n)$, then we can consider $x$ as
the natural number
\begin{equation}\label{madic}
x=\sum_{i=0}^{n}\alpha_i m^i.
\end{equation}
Moreover, this formula defines the one to one correspondence between
natural numbers (with zero) and the space of final sequences
$x=(\alpha_0,\alpha_1,\dots,\alpha_n)$.

We denote the space of sequences (\ref{K}), (\ref{K1}) by the symbol
$\mathbb{Z}_m$. Ultrametric is introduced on this set in the
following way. For two points
$$x=(\alpha_0, \alpha_1, \alpha_2,\dots, \alpha_n,\dots),\qquad
y=(\beta_0,\beta_1, \beta_2,\dots, \beta_n,\dots),
$$
we set
$$
d_m(x,y)= \frac{1}{m^k} \; \; \mbox{if}\; \;  \alpha_j= \beta_j,
j=0,1,\dots,k-1,\; \;  \mbox{and} \; \;\alpha_k\not=\beta_k.
$$
The ultrametric space $(X=\mathbb{Z}_m, d=d_m)$ is called the space
of $m$-adic integers.

The ultrametric $d_m$ describes the following hierarchical
structure. If $x=(\alpha_1, \alpha_2,\dots,\alpha_n,\dots)$,
$\alpha_j=0,1,\dots,m-1$, is a vector encoding information on some
object, then digits $\alpha_j$ have different weights. The digit
$\alpha_0$ is the most important, $\alpha_1$ dominates over
$\alpha_2,\dots, \alpha_n,\dots,$ and so on. Such hierarchic
information vectors can be created by living systems, e.g., by the
brain to process information. Applications of $m$-adic numbers to
information theory and, in particular, to description of cognitive
processes and complex social systems were developed in
\cite{Andr,Andr3,COGN1,COGN2,COGN3,COGN4,COGN5}.

\medskip

In applications we will use not only  ''one dimensional''  $m$-adic
spaces, but also cartesian products of a few spaces, e.g., $m$-adic
plane $\mathbb{Z}_m^2= \mathbb{Z}_m \times \mathbb{Z}_m$ and so on.
Our aim is to show that 2-adic plane structure was embedded in the
genetic code, see \cite{genetic_code}.

We remark that $m$-adic numbers for $m=p$, where $p$ is a prime
number were intensively used (during last 20 years) in mathematical
physics \cite{VVZ,obzor}. The number theoretic definition is as
follows.

Let us fix a prime number $p>1$. For example, fix $p=2$ or fix
$p=1999$. The example of ultrametric space is the field of $p$-adic
numbers $\mathbb{Q}_p,$ which is the completion of the field of
rational numbers with respect to $p$-adic norm $|x|_p$, defined as
follows: for a rational number $x=p^{\gamma}{m\over n},$ where
$\gamma=0, \pm 1, \pm 2,\dots,$ and $m$, $n$ are non-zero and are
not divisible by $p$, its $p$-adic norm is
$$
|x|_p=p^{-\gamma}.
$$
$p$-Adic norm is widely used in number theory and algebraic
geometry.

If a rational number is divisible by $p^\gamma$, where $\gamma$ is
very large, then its $p$-adic norm is very small. This hierarchy of
the degrees of $p$ (i.e. of values of the $p$--adic norm) gives the
hierarchical (ultrametric) structure of the $p$--adic norm. $p$-Adic
numbers are in one to one correspondence with the series
$$
x=\sum_{i=\gamma}^{\infty}x_{i}p^i,\qquad x_i=0,\dots,p-1,
$$
where $\gamma$ is integer and $x_{\gamma}\ne 0$ (this expansion is
the analogue of (\ref{madic})).

The unit ball in $\mathbb{Q}_p$ with center at $a=0$ zero coincides
with the space of $p$-adic integers $\mathbb{Z}_p$.

$p$-Adic numbers were actively used (since pioneer papers of
I.Volovich, \cite{Volovich}) in high energy physics (superstring
theory, cosmology) and theory of disordered systems (spin glasses),
see, e.g.,  review \cite{obzor} and the book \cite{VVZ}.

\section{2-Adic parametrization of the genetic code}

In the 2--adic parametrization approach of \cite{genetic_code} we
enumerate in some special way the set of 64 codons (triples of
nucleotides) by pairs of digits $(x,y)$, $x,y=0,1,2,\dots,7$. These
digits are in one to one correspondence with the triples
$(x_0,x_1,x_2)$ and $(y_0,y_1,y_2)$ of 0 and 1 (the expansions of
$x$ and $y$ over degrees of 2):
$$
x=x_0+2x_1+4x_2,\qquad y=y_0+2y_1+4y_2,\qquad x_i,y_i=0,1.
$$
2--Adic norm of $x$ is equal to $2^{-i}$, where $i$ is the number of
the first non zero $x_i$ in the above expansion (analogously for
$y$).

Each pair of digits $(x_i,y_i)$ is defined by a nucleotide using the
rule
$$
\begin{array}{|c|c|}\hline
A & G \cr\hline U & C \cr\hline
\end{array}=\begin{array}{|c|c|}\hline 00 & 01
\cr\hline 10 & 11 \cr\hline
\end{array}.
$$
Since the nucleotides $A=(0,0)$ and $G=(0,1)$ are purines, $U=(1,0)$
and $C=(1,1)$  are pyrimidines, the different first digits in the
above binary representation corresponds to the different chemical
types of the nucleotides. Namely, the nucleotide $(x,y)$ with $x=0$
is a purine, and the nucleotide $(x,y)$ with $x=1$ is a pyrimidine.

The second digit $y=0,1$ in the considered parametrization describes
the $H$--bonding character (weak for $y=0$ and strong for $y=1$).

Using the above correspondence between the nucleotides and the
digits 0 and 1, we introduce the correspondence between the codons
(triples of nucleotides) and the triples $(x_0,x_1,x_2)$ and
$(y_0,y_1,y_2)$ of 0 and 1 (equivalently, the corresponding
$x,y=0,1,2,\dots,7$) by the following prescription.

The second nucleotide in the codon defines the pair $(x_0,y_0)$, the
first nucleotide in the codon defines the pair $(x_1,y_1)$, and the
third nucleotide in the codon defines the pair $(x_2,y_2)$.

This rule is related to the following hierarchy of nucleotides in
the codon
$$
2>1>3
$$
i.e the second nucleotide in the codon is the most important (and
the largest in the 2--adic norm) and the third nucleotide is the
least important.

After that we make the special reshuffling of the values of $x$ and
$y$:
$$
0, 4, 2, 6,     1, 5,   3, 7 \mapsto 1,2,3,4,5,6,7,8.
$$
similar to made in \cite{ABK}.

Then we enumerate codons using the described above rule. Namely we
put the codons in the table $8\times 8$ with the natural 2--adic
norm (here the numbers of the lines and columns are $x$ and $y$
defined above):
$$
\begin{array}{|c c | c c|c c | c c|}

\hline AAA & AAG & GAA & GAG & AGA & AGG & GGA & GGG \cr

 AAU & AAC & GAU & GAC & AGU & AGC & GGU & GGC \cr

\hline UAA & UAG & CAA & CAG & UGA & UGG & CGA & CGG \cr

 UAU & UAC & CAU & CAC & UGU & UGC & CGU & CGC \cr

\hline AUA & AUG & GUA & GUG & ACA & ACG & GCA & GCG \cr

 AUU & AUC & GUU & GUC & ACU & ACC & GCU & GCC \cr

\hline UUA & UUG & CUA & CUG & UCA & UCG & CCA & CCG \cr

 UUU & UUC & CUU & CUC & UCU & UCC & CCU & CCC \cr

\hline\end{array}
$$

The $2\times 2$ quadrates here are 2--dimensional 2--adic balls (or
clusters) of the diameter $1/4$.

After application of the eucaryotic genetic (amino acid) code
(described in the Appendix) to the above table we get the table of
amino acids on the 2--adic plane
$$
\begin{array}{|c|c|c|c|}
\hline \begin{array}{c} {\rm K}  \cr \hline{\rm N}
\end{array}  & \begin{array}{c}
{\rm E}  \cr \hline{\rm D}
\end{array}  & \begin{array}{c}
{\rm R}  \cr \hline{\rm S}
\end{array}  & {\rm G}\cr

\hline\begin{array}{c} {\rm Ter}  \cr \hline{\rm Y}
\end{array}  & \begin{array}{c}
{\rm Q}  \cr \hline{\rm H}
\end{array}  & \begin{array}{c} {\rm Ter | W}  \cr
\hline{\rm C}
\end{array}  & {\rm R}\cr

\hline \begin{array}{c} {\rm I | M}  \cr \hline{\rm I}
\end{array}
 & {\rm V} & {\rm T} & {\rm A}    \cr
\hline
\begin{array}{c}
{\rm L}  \cr \hline{\rm F}
\end{array}

 & {\rm L}  & {\rm S}
 & {\rm P} \cr \hline
\end{array}
$$
where Ter is the stop codon. Each square in the above table is the 2
by 2 square in the 2-adic plane of codons. In particular, the
genetic code map acts as follows on the 2--adic balls
$$
\begin{array}{|c c|}\hline
AAA & AAG \cr AAU & AAC \cr\hline
\end{array}\to\begin{array}{|c|}\hline {\rm K}  \cr \hline{\rm N}\cr \hline
\end{array}\, ,\qquad
\begin{array}{|c c|}\hline
CCA & CCG \cr CCU & CCC \cr\hline
\end{array}\to\begin{array}{|c|}\hline {\rm P}\cr \hline
\end{array}
$$
Here we use the standard notations for the nucleotides A, U, G, C
and the amino acids.

We see, that the degeneracy of the genetic code in the above 2-adic
parametrization is described by the 2-adic proximity --- the codons
which encode the same amino acid are 2-adically close. Moreover, the
domains with the different degeneracy are symmetric at the 2-adic
plane --- the lower right half of the plane is occupied by amino
acids with the degeneracy four, and the upper left half of the plane
contains the amino acids with the degeneracy mainly equal to two. We
also have the five cases of additional degeneracy which is not
described by the 2-adic parametrization.

The described here 2-adic 2-dimensional parametrization of the
genetic code is related to physical--chemical properties of the
amino acids. Namely, the hydrophobic amino acids are clustered in
the following two 2-adic balls in the 2-adic plane:
$$
\begin{array}{|c|c|c|c|}
\hline \begin{array}{c} {\rm }  \cr \hline{\rm }
\end{array}  & \begin{array}{c}
{\rm }  \cr \hline{\rm }
\end{array}  & \begin{array}{c}
{\rm }  \cr \hline{\rm }
\end{array}  & {\rm }\cr

\hline\begin{array}{c} {\rm }  \cr \hline{\rm }
\end{array}  & \begin{array}{c}
{\rm }  \cr \hline{\rm }
\end{array}  & \begin{array}{c} {\rm Ter | W}  \cr
\hline{\rm C}
\end{array}  & {\rm }\cr

\hline \begin{array}{c} {\rm I | M}  \cr \hline{\rm I}
\end{array}
 & {\rm V} & {\rm } & {\rm }    \cr
\hline
\begin{array}{c}
{\rm L}  \cr \hline{\rm F}
\end{array}

 & {\rm L}  & {\rm }
 & {\rm } \cr \hline
\end{array}
$$

Here the hydrophobic amino acids are listed according to the book
\cite{FP}.

Using the 2-adic parametrization of the genetic code we can divide
all the set of amino acid in the groups $\{$K, N, E, D, Y, Q, H$\}$,
$\{$R, G, W, C$\}$, $\{$I, M, V, L, F$\}$, $\{$T, A, S, P$\}$. These
groups are the images with respect to the map of the genetic code of
the four quadrants (2--adic balls of the diameter $1/2$) of the
2--adic plane of codons.

\section{The PAM matrix}

The following table describes the PAM250 matrix:

$$
A=\begin{array}{|c|c c c c c c c c c c c  c c c c c  c c c c|}
\hline {\rm *} & {\rm A} & {\rm C} & {\rm D} & {\rm E} & {\rm F} &
{\rm G} & {\rm H} & {\rm I} & {\rm K} & {\rm L} & {\rm M} & {\rm N}
& {\rm P} & {\rm Q} & {\rm R} & {\rm S} & {\rm T} & {\rm V} & {\rm
W} & {\rm Y} \cr

\hline

{\rm A} & {\rm 2} & {\rm -2} & {\rm 0} & {\rm 0} & {\rm -4} & {\rm
1} & {\rm -1} & {\rm -1} & {\rm -1} & {\rm -2} & {\rm -1} & {\rm 0}
& {\rm 1} & {\rm 0} & {\rm -2} & {\rm 1} & {\rm 1} & {\rm 0} & {\rm
-6} & {\rm -3} \cr

{\rm C} & {\rm *} & {\rm 12} & {\rm -5} & {\rm -5} & {\rm -4} & {\rm
-3} & {\rm -3} & {\rm -2} & {\rm -5} & {\rm -6} & {\rm -5} & {\rm
-4} & {\rm -3} & {\rm -5} & {\rm -4} & {\rm 0} & {\rm -2} & {\rm -2}
& {\rm -8} & {\rm 0} \cr

{\rm D} & {\rm *} & {\rm *} & {\rm 4} & {\rm 3} & {\rm -6} & {\rm 1}
& {\rm 1} & {\rm -2} & {\rm 0} & {\rm -4} & {\rm -3} & {\rm 2} &
{\rm -1} & {\rm 2} & {\rm -1} & {\rm 0} & {\rm 0} & {\rm -2} & {\rm
-7} & {\rm -4} \cr

{\rm E} & {\rm *} & {\rm *} & {\rm *} & {\rm 4} & {\rm -5} & {\rm 0}
& {\rm 1} & {\rm -2} & {\rm 0} & {\rm -3} & {\rm -2} & {\rm 1} &
{\rm -1} & {\rm 2} & {\rm -1} & {\rm 0} & {\rm 0} & {\rm -2} & {\rm
-7} & {\rm -4} \cr

{\rm F} & {\rm *} & {\rm *} & {\rm *} & {\rm *} & {\rm 9} & {\rm -5}
& {\rm -2} & {\rm 1} & {\rm -5} & {\rm 2} & {\rm 0} & {\rm -4} &
{\rm -5} & {\rm -5} & {\rm -4} & {\rm -3} & {\rm -3} & {\rm -1} &
{\rm 0} & {\rm 7} \cr

{\rm G} & {\rm *} & {\rm *} & {\rm *} & {\rm *} & {\rm *} & {\rm 5}
& {\rm -2} & {\rm -3} & {\rm -2} & {\rm -4} & {\rm -3} & {\rm 0} &
{\rm -1} & {\rm -1} & {\rm -3} & {\rm 1} & {\rm 0} & {\rm -1} & {\rm
-7} & {\rm -5} \cr

{\rm H} & {\rm *} & {\rm *} & {\rm *} & {\rm *} & {\rm *} & {\rm *}
& {\rm 6} & {\rm -2} & {\rm 0} & {\rm -2} & {\rm -2} & {\rm 2} &
{\rm 0} & {\rm 3} & {\rm 2} & {\rm -1} & {\rm -1} & {\rm -2} & {\rm
-3} & {\rm 0} \cr

{\rm I} & {\rm *} & {\rm *} & {\rm *} & {\rm *} & {\rm *} & {\rm *}
& {\rm *} & {\rm 5} & {\rm -2} & {\rm 2} & {\rm 2} & {\rm -2} & {\rm
-2} & {\rm -2} & {\rm -2} & {\rm -1} & {\rm 0} & {\rm 4} & {\rm -5}
& {\rm -1} \cr

{\rm K} & {\rm *} & {\rm *} & {\rm *} & {\rm *} & {\rm *} & {\rm *}
& {\rm *} & {\rm *} & {\rm 5} & {\rm -3} & {\rm 0} & {\rm 1} & {\rm
-1} & {\rm 1} & {\rm 3} & {\rm 0} & {\rm 0} & {\rm -2} & {\rm -3} &
{\rm -4} \cr

{\rm L} & {\rm *} & {\rm *} & {\rm *} & {\rm *} & {\rm *} & {\rm *}
& {\rm *} & {\rm *} & {\rm *} & {\rm 6} & {\rm 4} & {\rm -3} & {\rm
-3} & {\rm -2} & {\rm -3} & {\rm -3} & {\rm -2} & {\rm 2} & {\rm -2}
& {\rm -1} \cr

{\rm M} & {\rm *} & {\rm *} & {\rm *} & {\rm *} & {\rm *} & {\rm *}
& {\rm *} & {\rm *} & {\rm *} & {\rm *} & {\rm 6} & {\rm -2} & {\rm
-2} & {\rm -1} & {\rm 0} & {\rm -2} & {\rm -1} & {\rm 2} & {\rm -4}
& {\rm -2} \cr

{\rm N} & {\rm *} & {\rm *} & {\rm *} & {\rm *} & {\rm *} & {\rm *}
& {\rm *} & {\rm *} & {\rm *} & {\rm *} & {\rm *} & {\rm 2} & {\rm
-1} & {\rm 1} & {\rm 0} & {\rm 1} & {\rm 0} & {\rm -2} & {\rm -4} &
{\rm -2} \cr

{\rm P} & {\rm *} & {\rm *} & {\rm *} & {\rm *} & {\rm *} & {\rm *}
& {\rm *} & {\rm *} & {\rm *} & {\rm *} & {\rm *} & {\rm *} & {\rm
6} & {\rm 0} & {\rm 0} & {\rm 1} & {\rm 0} & {\rm -1} & {\rm -6} &
{\rm -5} \cr

{\rm Q} & {\rm *} & {\rm *} & {\rm *} & {\rm *} & {\rm *} & {\rm *}
& {\rm *} & {\rm *} & {\rm *} & {\rm *} & {\rm *} & {\rm *} & {\rm
*} & {\rm 4} & {\rm 1} & {\rm -1} & {\rm -1} & {\rm -2} & {\rm -5} &
{\rm -4} \cr

{\rm R} & {\rm *} & {\rm *} & {\rm *} & {\rm *} & {\rm *} & {\rm *}
& {\rm *} & {\rm *} & {\rm *} & {\rm *} & {\rm *} & {\rm *} & {\rm
*} & {\rm *} & {\rm 6} & {\rm 0} & {\rm -1} & {\rm -2} & {\rm 2} &
{\rm -4} \cr

{\rm S} & {\rm *} & {\rm *} & {\rm *} & {\rm *} & {\rm *} & {\rm *}
& {\rm *} & {\rm *} & {\rm *} & {\rm *} & {\rm *} & {\rm *} & {\rm
*} & {\rm *} & {\rm *} & {\rm 2} & {\rm 1} & {\rm -1} & {\rm -2} &
{\rm -3} \cr

{\rm T} & {\rm *} & {\rm *} & {\rm *} & {\rm *} & {\rm *} & {\rm *}
& {\rm *} & {\rm *} & {\rm *} & {\rm *} & {\rm *} & {\rm *} & {\rm
*} & {\rm *} & {\rm *} & {\rm *} & {\rm 3} & {\rm 0} & {\rm -5} &
{\rm -3} \cr

{\rm V} & {\rm *} & {\rm *} & {\rm *} & {\rm *} & {\rm *} & {\rm *}
& {\rm *} & {\rm *} & {\rm *} & {\rm *} & {\rm *} & {\rm *} & {\rm
*} & {\rm *} & {\rm *} & {\rm *} & {\rm *} & {\rm 4} & {\rm -6} &
{\rm -2} \cr

{\rm W} & {\rm *} & {\rm *} & {\rm *} & {\rm *} & {\rm *} & {\rm *}
& {\rm *} & {\rm *} & {\rm *} & {\rm *} & {\rm *} & {\rm *} & {\rm
*} & {\rm *} & {\rm *} & {\rm *} & {\rm *} & {\rm *} & {\rm 17} &
{\rm 0} \cr

{\rm Y} & {\rm *} & {\rm *} & {\rm *} & {\rm *} & {\rm *} & {\rm *}
& {\rm *} & {\rm *} & {\rm *} & {\rm *} & {\rm *} & {\rm *} & {\rm
*} & {\rm *} & {\rm *} & {\rm *} & {\rm *} & {\rm *} & {\rm *} &
{\rm 10} \cr

\hline
\end{array}
$$

Because the matrix $A$ is symmetrical we put only the half of matrix
elements. This matrix looks irregular and has no any obvious
structure. In the next Section we will show that some reshuffling of
the numbers of the lines and columns will put this matrix in more
regular form.

\section{Reshuffling of matrix elements of the PAM matrix}

Let us enumerate the lines and the columns of the PAM matrix $A$ of
the previous Section using the 2--adic parametrization of the
genetic code. We get for $A$ the following:
$$
A=\begin{array}{|c|c c c c c c c| c c c c | c c c c c | c c c c|}
\hline {\rm *} & {\rm K} & {\rm N} & {\rm E} & {\rm D} & {\rm Y} &
{\rm Q} & {\rm H} & {\rm R} & {\rm G} & {\rm W} & {\rm C} & {\rm I}
& {\rm M} & {\rm V} & {\rm L} & {\rm F} & {\rm T} & {\rm A} & {\rm
S} & {\rm P} \cr

\hline

{\rm K} & {\rm 5} & {\rm 1} & {\rm 0} & {\rm 0} & {\rm -4} & {\rm 1}
& {\rm 0} & {\rm 3} & {\rm -2} & {\rm -3} & {\rm -5} & {\rm -2} &
{\rm 0} & {\rm -2} & {\rm -3} & {\rm -5} & {\rm 0} & {\rm -1} & {\rm
0} & {\rm -1} \cr

{\rm N} & {\rm 1} & {\rm 2} & {\rm 1} & {\rm 2} & {\rm -2} & {\rm 1}
& {\rm 2} & {\rm 0} & {\rm 0} & {\rm -4} & {\rm -4} & {\rm -2} &
{\rm -2} & {\rm -2} & {\rm -3} & {\rm -4} & {\rm 0} & {\rm 0} & {\rm
1} & {\rm -1} \cr

{\rm E} & {\rm 0} & {\rm 1} & {\rm 4} & {\rm 3} & {\rm -4} & {\rm 2}
& {\rm 1} & {\rm -1} & {\rm 0} & {\rm -7} & {\rm -5} & {\rm -2} &
{\rm -2} & {\rm -2} & {\rm -3} & {\rm -5} & {\rm 0} & {\rm 0} & {\rm
0} & {\rm -1} \cr

{\rm D} & {\rm 0} & {\rm 2} & {\rm 3} & {\rm 4} & {\rm -4} & {\rm 2}
& {\rm 1} & {\rm -1} & {\rm 1} & {\rm -7} & {\rm -5} & {\rm -2} &
{\rm -3} & {\rm -2} & {\rm -4} & {\rm -6} & {\rm 0} & {\rm 0} & {\rm
0} & {\rm -1} \cr

{\rm Y} & {\rm -4} & {\rm -2} & {\rm -4} & {\rm -4} & {\rm 10} &
{\rm -4} & {\rm 0} & {\rm -4} & {\rm -5} & {\rm 0} & {\rm 0} & {\rm
-1} & {\rm -2} & {\rm -2} & {\rm -1} & {\rm 7} & {\rm -3} & {\rm -3}
& {\rm -3} & {\rm -5} \cr

{\rm Q} & {\rm 1} & {\rm 1} & {\rm 2} & {\rm 2} & {\rm -4} & {\rm 4}
& {\rm 3} & {\rm 1} & {\rm -1} & {\rm -5} & {\rm -5} & {\rm -2} &
{\rm -1} & {\rm -2} & {\rm -2} & {\rm -5} & {\rm -1} & {\rm 0} &
{\rm -1} & {\rm 0} \cr

{\rm H} & {\rm 0} & {\rm 2} & {\rm 1} & {\rm 1} & {\rm 0} & {\rm 3}
& {\rm 6} & {\rm 2} & {\rm -2} & {\rm -3} & {\rm -3} & {\rm -2} &
{\rm -2} & {\rm -2} & {\rm -2} & {\rm -2} & {\rm -1} & {\rm -1} &
{\rm -1} & {\rm 0} \cr

\hline

{\rm R} & {\rm 3} & {\rm 0} & {\rm -1} & {\rm -1} & {\rm -4} & {\rm
1} & {\rm 2} & {\rm 6} & {\rm -3} & {\rm 2} & {\rm -4} & {\rm -2} &
{\rm 0} & {\rm -2} & {\rm -3} & {\rm -4} & {\rm -1} & {\rm -2} &
{\rm 0} & {\rm 0} \cr

{\rm G} & {\rm -2} & {\rm 0} & {\rm 0} & {\rm 1} & {\rm -5} & {\rm
-1} & {\rm -2} & {\rm -3} & {\rm 5} & {\rm -7} & {\rm -3} & {\rm -3}
& {\rm -3} & {\rm -1} & {\rm -4} & {\rm -5} & {\rm 0} & {\rm 1} &
{\rm 1} & {\rm -1} \cr

{\rm W} & {\rm -3} & {\rm -4} & {\rm -7} & {\rm -7} & {\rm 0} & {\rm
-5} & {\rm -3} & {\rm 2} & {\rm -7} & {\rm 17} & {\rm -8} & {\rm -5}
& {\rm -4} & {\rm -6} & {\rm -2} & {\rm 0} & {\rm -5} & {\rm -6} &
{\rm -2} & {\rm -6} \cr

{\rm C} & {\rm -5} & {\rm -4} & {\rm -5} & {\rm -5} & {\rm 0} & {\rm
-5} & {\rm -3} & {\rm -4} & {\rm -3} & {\rm -8} & {\rm 12} & {\rm
-2} & {\rm -5} & {\rm -2} & {\rm -6} & {\rm -4} & {\rm -2} & {\rm
-2} & {\rm 0} & {\rm -3} \cr

\hline

{\rm I} & {\rm -2} & {\rm -2} & {\rm -2} & {\rm -2} & {\rm -1} &
{\rm -2} & {\rm -2} & {\rm -2} & {\rm -3} & {\rm -5} & {\rm -2} &
{\rm 5} & {\rm 2} & {\rm 4} & {\rm 2} & {\rm 1} & {\rm 0} & {\rm -1}
& {\rm -1} & {\rm -2} \cr

{\rm M} & {\rm 0} & {\rm -2} & {\rm -2} & {\rm -3} & {\rm -2} & {\rm
-1} & {\rm -2} & {\rm 0} & {\rm -3} & {\rm -4} & {\rm -5} & {\rm 2}
& {\rm 6} & {\rm 2} & {\rm 4} & {\rm 0} & {\rm -1} & {\rm -1} & {\rm
-2} & {\rm -2} \cr

{\rm V} & {\rm -2} & {\rm -2} & {\rm -2} & {\rm -2} & {\rm -2} &
{\rm -2} & {\rm -2} & {\rm -2} & {\rm -1} & {\rm -6} & {\rm -2} &
{\rm 4} & {\rm 2} & {\rm 4} & {\rm 2} & {\rm -1} & {\rm 0} & {\rm 0}
& {\rm -1} & {\rm -1} \cr

{\rm L} & {\rm -3} & {\rm -3} & {\rm -3} & {\rm -4} & {\rm -1} &
{\rm -2} & {\rm -2} & {\rm -3} & {\rm -4} & {\rm -2} & {\rm -6} &
{\rm 2} & {\rm 4} & {\rm 2} & {\rm 6} & {\rm 2} & {\rm -2} & {\rm
-2} & {\rm -3} & {\rm -3} \cr

{\rm F} & {\rm -5} & {\rm -4} & {\rm -5} & {\rm -6} & {\rm 7} & {\rm
-5} & {\rm -2} & {\rm -4} & {\rm -5} & {\rm 0} & {\rm -4} & {\rm 1}
& {\rm 0} & {\rm -1} & {\rm 2} & {\rm 9} & {\rm -3} & {\rm -4} &
{\rm -3} & {\rm -5} \cr

\hline

{\rm T} & {\rm 0} & {\rm 0} & {\rm 0} & {\rm 0} & {\rm -3} & {\rm
-1} & {\rm -1} & {\rm -1} & {\rm 0} & {\rm -5} & {\rm -2} & {\rm 0}
& {\rm -1} & {\rm 0} & {\rm -2} & {\rm -3} & {\rm 3} & {\rm 1} &
{\rm 1} & {\rm 0} \cr

{\rm A} & {\rm -1} & {\rm 0} & {\rm 0} & {\rm 0} & {\rm -3} & {\rm
0} & {\rm -1} & {\rm -2} & {\rm 1} & {\rm -6} & {\rm -2} & {\rm -1}
& {\rm -1} & {\rm 0} & {\rm -2} & {\rm -4} & {\rm 1} & {\rm 2} &
{\rm 1} & {\rm 1} \cr

{\rm S} & {\rm 0} & {\rm 1} & {\rm 0} & {\rm 0} & {\rm -3} & {\rm
-1} & {\rm -1} & {\rm 0} & {\rm 1} & {\rm -2} & {\rm 0} & {\rm -1} &
{\rm -2} & {\rm -1} & {\rm -3} & {\rm -3} & {\rm 1} & {\rm 1} & {\rm
2} & {\rm 1} \cr

{\rm P} & {\rm -1} & {\rm -1} & {\rm -1} & {\rm -1} & {\rm -5} &
{\rm 0} & {\rm 0} & {\rm 0} & {\rm -1} & {\rm -6} & {\rm -3} & {\rm
-2} & {\rm -2} & {\rm -1} & {\rm -3} & {\rm -5} & {\rm 0} & {\rm 1}
& {\rm 1} & {\rm 6} \cr

\hline
\end{array}
$$

Here we divided all the set of amino acids to the groups  $\{$K, N,
E, D, Y, Q, H$\}$, $\{$R, G, W, C$\}$, $\{$I, M, V, L, F$\}$, $\{$T,
A, S, P$\}$. These four groups correspond to the second nucleotide
in the codons which encode (through the amino acid genetic code)
amino acids in the corresponding group. Namely, the first group
correspond to the codons with A (Adenine) at the second position,
the second group correspond to the codons with G (Guanine), the
third group correspond to the codons with U (Uracil), and the fourth
group correspond to the codons with C (Cytosine).

Compared to the PAM matrix from the previous Section, this
reshuffled matrix is more regular. It has large positive matrix
elements at the main diagonal, and off diagonal terms are close to
block constant at least at some of the 16 blocks, corresponding to
matrix elements with the indices from one of the 4 groups of amino
acids.

In particular, in the block corresponding to the matrix elements
$A_{ij}$, $i,j\in\{$T, A, S, P$\}$, all off diagonal matrix elements
are equal to 1 (10 matrix elements) or 0 (2 matrix elements).

In the block $A_{ij}$, $i\in\{$K, N, E, D, Y, Q, H$\}$, $j\in\{$T,
A, S, P$\}$ we have matrix elements equal to 0 (13 matrix elements),
$-1$ (10 matrix elements), 1 (1 matrix element) and anomalous matrix
elements equal to $-3$ and $-5$ corresponding to the amino acid Y.
The analogous situation we will have in the other blocks of the
matrix $A$.

We arrive to the following picture: the PAM matrix $A$ will be a
block matrix with matrix elements close to locally constant (i.e.
constant in the 16 blocks) if we will exclude matrix elements
corresponding to some amino acids, namely, to the amino acids Y, W,
C, L, F, and R.

\section{Expansion for the PAM matrix}

In the present Section we introduce the main construction of this
paper: we will expand the PAM matrix $A$ in the sum of the matrices
$A^{(2)}$ and $A^{(\infty)}$:
$$
A=A^{(2)}+A^{(\infty)},
$$
where the matrix $A^{(2)}$ will be 2--adically regular (matrix
elements are close to locally constant), and the matrix
$A^{(\infty)}$ will be sparse (i.e. majority of matrix elements of
this matrix will be equal to zero).

We propose the following choice for matrices $A^{(2)}$ and
$A^{(\infty)}$.
$$
A^{(2)}=\begin{array}{|c|c c c c c c c| c c c c | c c c c c | c c c
c|} \hline {\rm *} & {\rm K} & {\rm N} & {\rm E} & {\rm D} & {\rm Y}
& {\rm Q} & {\rm H} & {\rm R} & {\rm G} & {\rm W} & {\rm C} & {\rm
I} & {\rm M} & {\rm V} & {\rm L} & {\rm F} & {\rm T} & {\rm A} &
{\rm S} & {\rm P} \cr

\hline

{\rm K} & {\rm 5} & {\rm 1} & {\rm 1} & {\rm 1} & {\rm 1} & {\rm 1}
& {\rm 0} & {\rm -1} & {\rm -2} & {\rm -1} & {\rm -1} & {\rm -2} &
{\rm -2} & {\rm -2} & {\rm -2} & {\rm -2} & {\rm 0} & {\rm -1} &
{\rm 0} & {\rm -1} \cr

{\rm N} & {\rm 1} & {\rm 2} & {\rm 1} & {\rm 2} & {\rm 1} & {\rm 1}
& {\rm 2} & {\rm 0} & {\rm 0} & {\rm 0} & {\rm 0} & {\rm -2} & {\rm
-2} & {\rm -2} & {\rm -2} & {\rm -2} & {\rm 0} & {\rm 0} & {\rm 0} &
{\rm -1} \cr

{\rm E} & {\rm 1} & {\rm 1} & {\rm 4} & {\rm 3} & {\rm 1} & {\rm 2}
& {\rm 1} & {\rm -1} & {\rm 0} & {\rm -1} & {\rm -1} & {\rm -2} &
{\rm -2} & {\rm -2} & {\rm -2} & {\rm -2} & {\rm 0} & {\rm 0} & {\rm
0} & {\rm -1} \cr

{\rm D} & {\rm 1} & {\rm 2} & {\rm 3} & {\rm 4} & {\rm 1} & {\rm 2}
& {\rm 1} & {\rm -1} & {\rm 0} & {\rm -1} & {\rm -1} & {\rm -2} &
{\rm -3} & {\rm -2} & {\rm -2} & {\rm -2} & {\rm 0} & {\rm 0} & {\rm
0} & {\rm -1} \cr

{\rm Y} & {\rm 1} & {\rm 1} & {\rm 1} & {\rm 1} & {\rm 10} & {\rm 1}
& {\rm 0} & {\rm -1} & {\rm -1} & {\rm 0} & {\rm 0} & {\rm -1} &
{\rm -2} & {\rm -2} & {\rm -1} & {\rm -2} & {\rm -1} & {\rm -1} &
{\rm -1} & {\rm -1} \cr

{\rm Q} & {\rm 1} & {\rm 1} & {\rm 2} & {\rm 2} & {\rm 1} & {\rm 4}
& {\rm 3} & {\rm -1} & {\rm -1} & {\rm -1} & {\rm -1} & {\rm -2} &
{\rm -1} & {\rm -2} & {\rm -2} & {\rm -2} & {\rm -1} & {\rm 0} &
{\rm -1} & {\rm 0} \cr

{\rm H} & {\rm 0} & {\rm 2} & {\rm 1} & {\rm 1} & {\rm 0} & {\rm 3}
& {\rm 6} & {\rm -1} & {\rm -2} & {\rm -1} & {\rm -1} & {\rm -2} &
{\rm -2} & {\rm -2} & {\rm -2} & {\rm -2} & {\rm -1} & {\rm -1} &
{\rm -1} & {\rm 0} \cr

\hline

{\rm R} & {\rm -1} & {\rm 0} & {\rm -1} & {\rm -1} & {\rm -1} & {\rm
-1} & {\rm -1} & {\rm 6} & {\rm -3} & {\rm -2} & {\rm -2} & {\rm -2}
& {\rm -2} & {\rm -2} & {\rm -2} & {\rm -2} & {\rm -1} & {\rm -2} &
{\rm 0} & {\rm 0} \cr

{\rm G} & {\rm -2} & {\rm 0} & {\rm 0} & {\rm 0} & {\rm -1} & {\rm
-1} & {\rm -2} & {\rm -3} & {\rm 5} & {\rm -3} & {\rm -3} & {\rm -3}
& {\rm -3} & {\rm -1} & {\rm -2} & {\rm -3} & {\rm 0} & {\rm -1} &
{\rm 1} & {\rm -1} \cr

{\rm W} & {\rm -1} & {\rm 0} & {\rm -1} & {\rm -1} & {\rm 0} & {\rm
-1} & {\rm -1} & {\rm -2} & {\rm -3} & {\rm 17} & {\rm -3} & {\rm
-2} & {\rm -2} & {\rm -2} & {\rm -2} & {\rm -2} & {\rm -2} & {\rm
-2} & {\rm -2} & {\rm -2} \cr

{\rm C} & {\rm -1} & {\rm 0} & {\rm -1} & {\rm -1} & {\rm 0} & {\rm
-1} & {\rm -1} & {\rm -2} & {\rm -3} & {\rm -3} & {\rm 12} & {\rm
-2} & {\rm -2} & {\rm -2} & {\rm -2} & {\rm -2} & {\rm -2} & {\rm
-2} & {\rm -2} & {\rm -3} \cr

\hline

{\rm I} & {\rm -2} & {\rm -2} & {\rm -2} & {\rm -2} & {\rm -1} &
{\rm -2} & {\rm -2} & {\rm -2} & {\rm -3} & {\rm -2} & {\rm -2} &
{\rm 5} & {\rm 2} & {\rm 2} & {\rm 2} & {\rm 2} & {\rm 0} & {\rm -1}
& {\rm -1} & {\rm -2} \cr

{\rm M} & {\rm -2} & {\rm -2} & {\rm -2} & {\rm -3} & {\rm -2} &
{\rm -1} & {\rm -2} & {\rm -2} & {\rm -3} & {\rm -2} & {\rm -2} &
{\rm 2} & {\rm 6} & {\rm 2} & {\rm 2} & {\rm 2} & {\rm -1} & {\rm
-1} & {\rm -2} & {\rm -2} \cr

{\rm V} & {\rm -2} & {\rm -2} & {\rm -2} & {\rm -2} & {\rm -2} &
{\rm -2} & {\rm -2} & {\rm -2} & {\rm -1} & {\rm -2} & {\rm -2} &
{\rm 2} & {\rm 2} & {\rm 4} & {\rm 2} & {\rm 2} & {\rm 0} & {\rm 0}
& {\rm -1} & {\rm -1} \cr

{\rm L} & {\rm -2} & {\rm -2} & {\rm -2} & {\rm -2} & {\rm -1} &
{\rm -2} & {\rm -2} & {\rm -2} & {\rm -3} & {\rm -2} & {\rm -2} &
{\rm 2} & {\rm 2} & {\rm 2} & {\rm 6} & {\rm 2} & {\rm -1} & {\rm
-1} & {\rm -2} & {\rm -2} \cr

{\rm F} & {\rm -2} & {\rm -2} & {\rm -2} & {\rm -2} & {\rm -2} &
{\rm -2} & {\rm -2} & {\rm -2} & {\rm -3} & {\rm -2} & {\rm -2} &
{\rm 2} & {\rm 2} & {\rm 2} & {\rm 2} & {\rm 9} & {\rm -1} & {\rm
-1} & {\rm -2} & {\rm -2} \cr

\hline

{\rm T} & {\rm 0} & {\rm 0} & {\rm 0} & {\rm 0} & {\rm -1} & {\rm
-1} & {\rm -1} & {\rm -1} & {\rm 0} & {\rm -2} & {\rm -2} & {\rm 0}
& {\rm -1} & {\rm 0} & {\rm -1} & {\rm -1} & {\rm 3} & {\rm 1} &
{\rm 1} & {\rm 0} \cr

{\rm A} & {\rm -1} & {\rm 0} & {\rm 0} & {\rm 0} & {\rm -1} & {\rm
0} & {\rm -1} & {\rm -2} & {\rm -1} & {\rm -2} & {\rm -2} & {\rm -1}
& {\rm -1} & {\rm 0} & {\rm -1} & {\rm -1} & {\rm 1} & {\rm 2} &
{\rm 1} & {\rm 1} \cr

{\rm S} & {\rm 0} & {\rm 0} & {\rm 0} & {\rm 0} & {\rm -1} & {\rm
-1} & {\rm -1} & {\rm 0} & {\rm 1} & {\rm -2} & {\rm -2} & {\rm -1}
& {\rm -2} & {\rm -1} & {\rm -2} & {\rm -2} & {\rm 1} & {\rm 1} &
{\rm 2} & {\rm 1} \cr

{\rm P} & {\rm -1} & {\rm -1} & {\rm -1} & {\rm -1} & {\rm -1} &
{\rm 0} & {\rm 0} & {\rm 0} & {\rm -1} & {\rm -2} & {\rm -3} & {\rm
-2} & {\rm -2} & {\rm -1} & {\rm -2} & {\rm -2} & {\rm 0} & {\rm 1}
& {\rm 1} & {\rm 6} \cr

\hline
\end{array}
$$

$$
A^{(\infty)}=\begin{array}{|c|c c c c c c c| c c c c | c c c c c | c
c c c|} \hline {\rm *} & {\rm K} & {\rm N} & {\rm E} & {\rm D} &
{\rm Y} & {\rm Q} & {\rm H} & {\rm R} & {\rm G} & {\rm W} & {\rm C}
& {\rm I} & {\rm M} & {\rm V} & {\rm L} & {\rm F} & {\rm T} & {\rm
A} & {\rm S} & {\rm P} \cr

\hline

{\rm K} & {\rm 0} & {\rm 0} & {\rm -1} & {\rm -1} & {\rm -5} & {\rm
0} & {\rm 0} & {\rm 4} & {\rm 0} & {\rm -2} & {\rm -4} & {\rm 0} &
{\rm 2} & {\rm 0} & {\rm -1} & {\rm -3} & {\rm 0} & {\rm 0} & {\rm
0} & {\rm 0} \cr

{\rm N} & {\rm 0} & {\rm 0} & {\rm 0} & {\rm 0} & {\rm -3} & {\rm 0}
& {\rm 0} & {\rm 0} & {\rm 0} & {\rm -4} & {\rm -4} & {\rm 0} & {\rm
0} & {\rm 0} & {\rm -1} & {\rm -2} & {\rm 0} & {\rm 0} & {\rm 1} &
{\rm 0} \cr

{\rm E} & {\rm -1} & {\rm 0} & {\rm 0} & {\rm 0} & {\rm -5} & {\rm
0} & {\rm 0} & {\rm 0} & {\rm 0} & {\rm -6} & {\rm -4} & {\rm 0} &
{\rm 0} & {\rm 0} & {\rm -1} & {\rm -3} & {\rm 0} & {\rm 0} & {\rm
0} & {\rm 0} \cr

{\rm D} & {\rm -1} & {\rm 0} & {\rm 0} & {\rm 0} & {\rm -5} & {\rm
0} & {\rm 0} & {\rm 0} & {\rm 1} & {\rm -6} & {\rm -4} & {\rm 0} &
{\rm 0} & {\rm 0} & {\rm -2} & {\rm -4} & {\rm 0} & {\rm 0} & {\rm
0} & {\rm 0} \cr

{\rm Y} & {\rm -5} & {\rm -3} & {\rm -5} & {\rm -5} & {\rm 0} & {\rm
-5} & {\rm 0} & {\rm -3} & {\rm -4} & {\rm 0} & {\rm 0} & {\rm 0} &
{\rm 0} & {\rm 0} & {\rm 0} & {\rm 9} & {\rm -2} & {\rm -2} & {\rm
-2} & {\rm -4} \cr

{\rm Q} & {\rm 0} & {\rm 0} & {\rm 0} & {\rm 0} & {\rm -5} & {\rm 0}
& {\rm 0} & {\rm 2} & {\rm 0} & {\rm -4} & {\rm -4} & {\rm 0} & {\rm
0} & {\rm 0} & {\rm 0} & {\rm 0} & {\rm 0} & {\rm 0} & {\rm 0} &
{\rm 0} \cr

{\rm H} & {\rm 0} & {\rm 0} & {\rm 0} & {\rm 0} & {\rm 0} & {\rm 0}
& {\rm 0} & {\rm 3} & {\rm 0} & {\rm -2} & {\rm -2} & {\rm 0} & {\rm
0} & {\rm 0} & {\rm 0} & {\rm 0} & {\rm 0} & {\rm 0} & {\rm 0} &
{\rm 0} \cr

\hline

{\rm R} & {\rm 4} & {\rm 0} & {\rm 0} & {\rm 0} & {\rm -3} & {\rm 2}
& {\rm 3} & {\rm 0} & {\rm 0} & {\rm 4} & {\rm -2} & {\rm 0} & {\rm
2} & {\rm 0} & {\rm -1} & {\rm -2} & {\rm 0} & {\rm 0} & {\rm 0} &
{\rm 0} \cr

{\rm G} & {\rm 0} & {\rm 0} & {\rm 0} & {\rm 1} & {\rm -4} & {\rm 0}
& {\rm 0} & {\rm 0} & {\rm 0} & {\rm -4} & {\rm 0} & {\rm 0} & {\rm
0} & {\rm 0} & {\rm -1} & {\rm -2} & {\rm 0} & {\rm 2} & {\rm 0} &
{\rm 0} \cr

{\rm W} & {\rm -2} & {\rm -4} & {\rm -6} & {\rm -6} & {\rm 0} & {\rm
-4} & {\rm -2} & {\rm 4} & {\rm -4} & {\rm 0} & {\rm -5} & {\rm -3}
& {\rm -2} & {\rm -4} & {\rm 0} & {\rm 2} & {\rm -3} & {\rm -4} &
{\rm 0} & {\rm -4} \cr

{\rm C} & {\rm -4} & {\rm -4} & {\rm -4} & {\rm -4} & {\rm 0} & {\rm
-4} & {\rm -2} & {\rm -2} & {\rm 0} & {\rm -5} & {\rm 0} & {\rm 0} &
{\rm -3} & {\rm 0} & {\rm -4} & {\rm -2} & {\rm 0} & {\rm 0} & {\rm
2} & {\rm 0} \cr

\hline

{\rm I} & {\rm 0} & {\rm 0} & {\rm 0} & {\rm 0} & {\rm 0} & {\rm 0}
& {\rm 0} & {\rm 0} & {\rm 0} & {\rm -3} & {\rm 0} & {\rm 0} & {\rm
0} & {\rm 2} & {\rm 0} & {\rm -1} & {\rm 0} & {\rm 0} & {\rm 0} &
{\rm 0} \cr

{\rm M} & {\rm 2} & {\rm 0} & {\rm 0} & {\rm 0} & {\rm 0} & {\rm 0}
& {\rm 0} & {\rm 2} & {\rm 0} & {\rm -2} & {\rm -3} & {\rm 0} & {\rm
0} & {\rm 0} & {\rm 2} & {\rm -2} & {\rm 0} & {\rm 0} & {\rm 0} &
{\rm 0} \cr

{\rm V} & {\rm 0} & {\rm 0} & {\rm 0} & {\rm 0} & {\rm 0} & {\rm 0}
& {\rm 0} & {\rm 0} & {\rm 0} & {\rm -4} & {\rm 0} & {\rm 2} & {\rm
0} & {\rm 0} & {\rm 0} & {\rm -3} & {\rm 0} & {\rm 0} & {\rm 0} &
{\rm 0} \cr

{\rm L} & {\rm -1} & {\rm -1} & {\rm -1} & {\rm -2} & {\rm 0} & {\rm
0} & {\rm 0} & {\rm -1} & {\rm -1} & {\rm 0} & {\rm -4} & {\rm 0} &
{\rm 2} & {\rm 0} & {\rm 0} & {\rm 0} & {\rm -1} & {\rm -1} & {\rm
-1} & {\rm -1} \cr

{\rm F} & {\rm -3} & {\rm -2} & {\rm -3} & {\rm -4} & {\rm 9} & {\rm
-3} & {\rm 0} & {\rm -2} & {\rm -2} & {\rm 2} & {\rm -2} & {\rm -1}
& {\rm -2} & {\rm -3} & {\rm 0} & {\rm 0} & {\rm -2} & {\rm -3} &
{\rm -1} & {\rm -3} \cr

\hline

{\rm T} & {\rm 0} & {\rm 0} & {\rm 0} & {\rm 0} & {\rm -2} & {\rm 0}
& {\rm 0} & {\rm 0} & {\rm 0} & {\rm -3} & {\rm 0} & {\rm 0} & {\rm
0} & {\rm 0} & {\rm -1} & {\rm -2} & {\rm 0} & {\rm 0} & {\rm 0} &
{\rm 0} \cr

{\rm A} & {\rm 0} & {\rm 0} & {\rm 0} & {\rm 0} & {\rm -2} & {\rm 0}
& {\rm 0} & {\rm 0} & {\rm 2} & {\rm -4} & {\rm 0} & {\rm 0} & {\rm
0} & {\rm 0} & {\rm -1} & {\rm -3} & {\rm 0} & {\rm 0} & {\rm 0} &
{\rm 0} \cr

{\rm S} & {\rm 0} & {\rm 1} & {\rm 0} & {\rm 0} & {\rm -2} & {\rm 0}
& {\rm 0} & {\rm 0} & {\rm 0} & {\rm 0} & {\rm 2} & {\rm 0} & {\rm
0} & {\rm 0} & {\rm -1} & {\rm -1} & {\rm 0} & {\rm 0} & {\rm 0} &
{\rm 0} \cr

{\rm P} & {\rm 0} & {\rm 0} & {\rm 0} & {\rm 0} & {\rm -4} & {\rm 0}
& {\rm 0} & {\rm 0} & {\rm 0} & {\rm -4} & {\rm 0} & {\rm 0} & {\rm
0} & {\rm 0} & {\rm -1} & {\rm -3} & {\rm 0} & {\rm 0} & {\rm 0} &
{\rm 0} \cr

\hline
\end{array}
$$

Non zero matrix elements of $A^{(\infty)}$ are mainly concentrated
on the lines and columns corresponding to Y, W, C, L, F. There are
also several non--zero matrix elements corresponding to R and some
other amino acids.

We see that the non zero matrix elements of $A^{(\infty)}$ are
mainly concentrated on aromatic amino acids, such as Y, F, W, and on
C which contains the SH group. Therefore deviations from 2--adic
regularity (i.e. the block structure of the $A^{(2)}$ matrix) can be
discussed as related to the geometric properties of the side chains
of amino acids (for aromatic amino acids Y, F, W, and for Arginine
R), and to the ability of Cysteine C to create a disulfide bond.

\begin{figure}
\begin{center}
\includegraphics[scale=0.7]{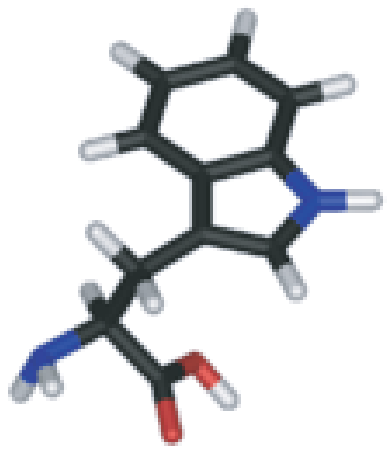}
\includegraphics[scale=0.7]{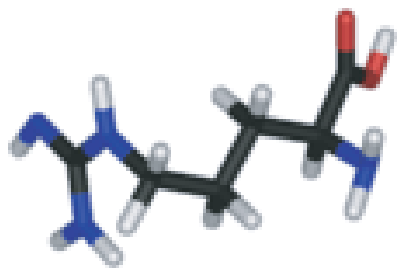}
\includegraphics[scale=0.4]{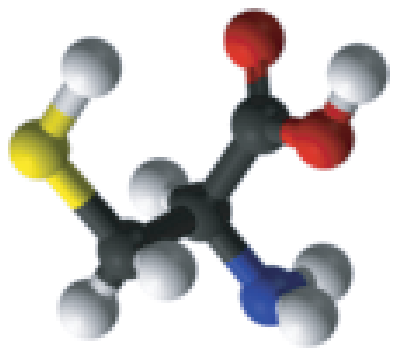}
\end{center}
\caption{tryptophan W, arginine R, cysteine C}\label{fig1}
\begin{center}
\includegraphics[scale=0.7]{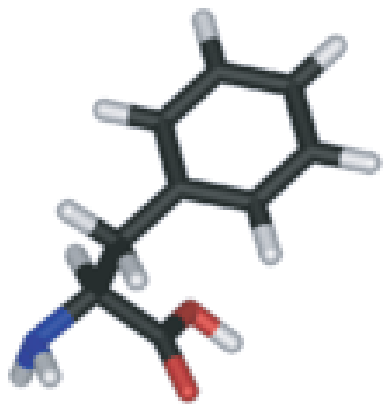}
\includegraphics[scale=0.7]{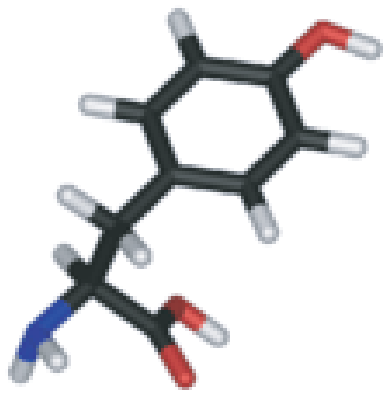}
\end{center}
\caption{phenylalanine F, tyrosine Y} \label{fig2}
\end{figure}

The above pictures show the amino acids W, R, C, F, Y corresponding
to non-zero matrix elements of the matrix $A^{(\infty)}$. Let us
mention that amino acids F and Y for which the corresponding matrix
element $A^{(\infty)}_{FY}$ is very large (and majority of the other
matrix elements of $A^{(\infty)}$  for these amino acids are
negative) are very similar from the point of view of geometry.

Of course, our analysis of PAM matrix based on the 2-adic plane
representation of the genetic code is only the first step in using
$p$-adic numbers in genetics and bioinformatics in general. We hope
to proceed towards other important problems, cf., e.g., \cite{PAM}--
\cite{TTT1}, \cite{TTT2}.  Finally, we mention the famous "In Silico
Biology" project, see, e.g., Yamato et al. \cite{TTT3}. We point out
that, in fact, operation of any computer can be represented as
2-adic dynamical system, see \cite{Andr3}, \cite{ANK}. Therefore
2-adic representation of the genetic code might be useful in
realization of the "In Silico Biology" project.

\section{Appendix: the PAM matrix}
\label{PAM}

In this section we discuss the construction of the Dayhoff PAM
matrix, which can be found for example in \cite{PAM,Isaev}. We start
with blocks --- ungapped multiple alignments of proteins from
existing databases. Any sequence in the block is no more than $15\%$
different from any other sequence in this block.

Then, the Markov model, which reproduces the mentioned blocks of
proteins was constructed. This Markov model is defined by the amino
acid substitutions (point mutations). We have the stationary
distribution $p_a$ of the probabilities of the amino acids,
$\sum_{a=1}^{20}p_a=1$, and the transition probability $p_{ab}$,
normalized by the condition that the probability of a point mutation
(substitution of the amino acid) at one step of the Markov model is
equal to $0.01$:
$$
\sum_{a,b=1}^{20}p_{ab}p_b=0.01.
$$

Then we take the matrix given by the $n$ steps of the Markov model,
i.e. the $n$-th degree $P^n$ of the matrix $P=(p_{ab})$, and
consider the matrix with the matrix elements
$$
A^{(n)}=\log_{10}\left({(P^n)_{ab}\over p_b}\right).
$$
This matrix is known as the PAM matrix (usually $n$ is taken to be
equal to 250 and the matrix elements are approximated by integers).

\section{Appendix: the genetic code}

The following table describes the eucaryotic genetic code --- the
correspondence between codons (triples of nucleotides and amino
acids):

\medskip

\noindent
\begin{tabular}{|c|c|c|c|}
 \hline \  & \  &  \ & \ \\
  AAA ~K  &  UAA ~Ter &  GAA ~E &  CAA ~Q  \\
  AAU ~N  &  UAU ~~~Y &  GAU ~D &  CAU ~H  \\
  AAG ~K  &  UAG ~Ter &  GAG ~E &  CAG ~Q  \\
  AAC ~N  &  UAC ~~~Y &  GAC ~D &  CAC ~H  \\
 \hline \  & \  & \  &   \\
  AUA ~I  &  UUA ~L &  GUA ~V &  CUA ~L \\
  AUU ~I  &  UUU ~F &  GUU ~V &  CUU ~L \\
  AUG ~M  &  UUG ~L &  GUG ~V &  CUG ~L \\
  AUC ~I  &  UUC ~F &  GUC ~V &  CUC ~L \\
 \hline \ & \   & \  &   \\
  AGA ~R  &  UGA ~Ter &  GGA ~G &  CGA ~R  \\
  AGU ~S  &  UGU ~~~C &  GGU ~G &  CGU ~R  \\
  AGG ~R  &  UGG ~~~W &  GGG ~G &  CGG ~R  \\
  AGC ~S  &  UGC ~~~C &  GGC ~G &  CGC ~R  \\
 \hline \ & \ & \ & \\
  ACA ~T  &  UCA ~S &  GCA ~A &  CCA ~P  \\
  ACU ~T  &  UCU ~S &  GCU ~A &  CCU ~P  \\
  ACG ~T  &  UCG ~S &  GCG ~A &  CCG ~P  \\
  ACC ~T  &  UCC ~S &  GCC ~A &  CCC ~P  \\
\hline
\end{tabular}

\bigskip\bigskip

\noindent{\bf Acknowledgments}\qquad The authors would like to thank
B.Dragovich, I.V.Volovich for fruitful discussions and valuable
comments. The authors were partially supported by the grant of
International center for mathematical modeling in physics,
engineering and cognitive science, University of Vaxjo. One of the
authors (A.K.) was partially supported by the QPIC grant, Tokyo
University of Science. One of the authors (S.K.) gratefully
acknowledges being partially supported by the grants DFG Project 436
RUS 113/809/0-1 and DFG Project 436 RUS 113/951, by the grants of
The Russian Foundation for Basic Research RFFI 05-01-04002-NNIO-a
and RFFI 08-01-00727-a, by the grant of the President of Russian
Federation for the support of scientific schools NSh-3224.2008.1, by
the Program of the Department of Mathematics of Russian Academy of
Science ''Modern problems of theoretical mathematics'' and by the
program of Ministry of Education and Science of Russia ''Development
of the scientific potential of High School, years of 2009--2010'',
project 3341.


\begin{thebibliography}{99}

\bibitem{PAM}  R. Durbin, S.R. Eddy, A. Krogh, G. Mitchison, {\it Biological Sequence Analysis: Probabilistic Models of Proteins and Nucleic
Acids}, Cambridge University Press, 1998.

\bibitem{Isaev} A.Isaev, {\it Introduction to mathematical methods in
bioinformatics}, Springer, 2006.

\bibitem{TTT} M. Nishihama, Yu. Sakatsuji, A. Arinami, and S. Miyazaki,
Informational approach for the study of cis-regulatory
elements and DNA binding proteins. In: L. Accardi, W. Freudenberg, M. Ohya (eds.), Quantum Bio-Informatics,
pp. 371-- 380. WSP, Singapore, 2007.

\bibitem{TTT1}  T. Suzuki and S. Miyazaki,
Basics of genome sequence analysis in bioinformatics - its fundamental ideas and problems
In: L. Accardi, W. Freudenberg, M. Ohya (eds.), Quantum Bio-Informatics,
pp. 299 -- 313. WSP, Singapore, 2008.


\bibitem{Murtagh} F.Murtagh, A.Heck, {\it Multivariate Data Analysis}, Kluwer Academic Publishers, Dordrecht, 1987.

\bibitem{VVZ} V.S. Vladimirov,  I.V. Volovich, Ye.I. Zelenov,
\emph{$p$--Adic analysis and mathematical physics}, World
Scientific, Singapore,  1994 (See also  Nauka, Moscow, 1994, in
Russian).

\bibitem{Andr}  A. Khrennikov, {\it Non--Archimedean Analysis:
Quantum Paradoxes, Dynamical Systems and Biological Models}, Kluwer
Academic Publishers, 1997.

\bibitem{Kozyrev} S.V. Kozyrev, {\it Methods and applications of ultrametric and $p$--adic analysis:
from wavelet theory to biophysics}. Modern problems of mathematics.
Issue 12. Steklov Mathematical Institute, Moscow, 2008, (in Russian)
http://www.mi.ras.ru/spm/pdf/012.pdf.


\bibitem{obzor} B. Dragovich, A. Yu. Khrennikov, S. V. Kozyrev and I. V.
Volovich, On $p$-adic mathematical physics.  $p$-Adic Numbers,
Ultrametric Analysis and Applications, {\bf 1}, N 1, 1-17 (2009).

\bibitem{Andr3} A.Yu. Khrennikov, {\it Information dynamics in cognitive, psychological and
anomalous phenomena}, Series in Fundamental Theories of Physics,
Kluwer, Dordrecht,  2004.


\bibitem{genetic_code} A.Yu. Khrennikov, S.V. Kozyrev, Genetic code on the diadic
plane //  Physica A: Statistical Mechanics and its Applications.
2007. V.381. P.265-272. arXiv:q-bio.QM/0701007

\bibitem{DD} B.Dragovich, A.Dragovich,
A $p$-Adic Model of DNA Sequence and Genetic Code, $p$-Adic Numbers,
Ultrametric Analysis and Applications, {\bf 1}, N 1, 34-41 (2009).
arXiv:q-bio/0607018v1

\bibitem{Andr5} A.Yu. Khrennikov, $p$--Adic information space and
gene expression. In: Integrative approaches to brain complexity,
eds. S.Grant, N.Heintz, J.Noebels, Welcome Truct Publ. P.14. 2006.

\bibitem{Swanson} R.Swanson, A unifying concept for the amino acid
code, Bulletin of Mathematical Biology, 1984. V.46. N.2. P.187-203.


\bibitem{SjoWold} M.Sj\"ostrom, S.Wold, A multivariate study of the
relationship between the genetic code and the physical--chemical
properties of amino acids. Journal of Molecular Evolution. 1985.
V.22. P.272-277.

\bibitem{Pattern} M.D.Perlwitz, C.Burks, M.S.Waterman, Pattern
Analysis of the Genetic Code, Advances in applied mathematics, 1988.
V.9. P.7-21.

\bibitem{ABK} V.A.Avetisov, A.H.Bikulov, S.V.Kozyrev,
Application of $p$--adic analysis to models of spontaneous breaking
of replica symmetry, // J. Phys. A: Math. Gen. 1999. V.32. N.50.
P.8785--8791, arXiv:cond-mat/9904360

\bibitem{PaSu}  G.Parisi,  N.Sourlas, $p$--Adic numbers and replica symmetry
breaking // European Phys. J. B. 2000.  V.14. P.535--542.
arXiv:cond-mat/9906095

\bibitem{COGN1} A. Yu. Khrennikov, Probabilistic pathway representation of
cognitive information. {\it J. Theor. Biology,} {\bf 231}, 597-613
(2004).

\bibitem{COGN2} A. Yu. Khrennikov, $p$-adic discrete dynamical systems and
collective behaviour of information states in cognitive models. {\it
Discrete Dynamics in Nature and Society,} {\bf 5,} 59-69 (2000).

\bibitem{COGN3} S.Albeverio,  A.Yu.Khrennikov, P.Kloeden, Memory retrieval as a
$p$-adic dynamical system. Biosystems, 49, 105-115 (1999).

\bibitem{COGN4} D.Dubischar, V.M.Gundlach, O.Steinkamp, A. Yu.Khrennikov, A
$p$-adic model for the process of thinking disturbed by
physiological and information noise. J. Theor. Biology,197, 451-467
(1999).

\bibitem{COGN5} A.Yu. Khrennikov, Human subconscious as the $p$-adic dynamical
system. J. of Theor. Biology. 193, 179-196 (1998).

\bibitem{Volovich} I.V.Volovich, $p$-Adic string, Class. Quantum
Gravity. 1987. V.4. L.83-L87. \\
I.V.Volovich, Number theory as the ultimate physical theory.
Preprint No. TH 4781/87, CERN, Geneva, 1987.

\bibitem{FP} A.V.Finkelshtein, O.B.Ptitsyn, {\it Physics of Proteins}, Academic Press,
London, 2002.

\bibitem{TTT2} D. Wanke, J. Killan, A basic introduction to gene expression studies using
microarray expression data analysis.
In: L. Accardi, W. Freudenberg, M. Ohya (eds.), Quantum Bio-Informatics,
pp. 314 -- 326. WSP, Singapore, 2008.

\bibitem{TTT3} I. Yamato, T. Ando, A. Suzuki, K. Harada, S. Itoh, S. Miyazaki, N. Kobayashi, M. Takeda,
Toward In Silico Biology (from sequences to systems).
In: L. Accardi, W. Freudenberg, M. Ohya (eds.), Quantum Bio-Informatics,
pp. 440 -- 455. WSP, Singapore, 2007.

\bibitem{ANK} V. Anashin and A. Yu. Khrennikov, {\it Applied algebraic
dynamics.} De Gruyter, Berlin (2009).


\end{thebibliography}
\end{document}